\documentclass[12pt,fleqn]{article}
\usepackage{amssymb}

\usepackage{amsmath}

\usepackage{amsmath,amssymb,amsthm}

\begin{document}
\begin{center}
{\Large \textbf{General Solution for a Coupled System\\ of Eikonal Equations in Two Space Variables}}

\vskip 20pt {\large \textbf{Irina YEHORCHENKO}}

\vskip 20pt {Institute of Mathematics of NAS Ukraine, 3 Tereshchenkivs'ka Str., Kyiv-4, Ukraine} \\
E-mail: iyegorch@imath.kiev.ua
\end{center}

\begin{abstract}
A general solution for a coupled system of eikonal equations
\begin{gather*}
u_\mu u_\mu =0,
\\
v_\mu v_\mu =0,
\\
u_\mu v_\mu =1
\end{gather*}
is presented, where lower indices designate derivatives, $\mu=0,1,2$ and summation is implied over the repeated indices.

This solution is of interest by itself due to wide applications of the eikonal equations, but the system considered also appears to be part of the reduction conditions for many equations of mathematical physics.

We describe in detail the procedure that allowed obtaining of the general solution using hodograph and contact transformations of the initial system; however, we omit here special case when the system is equivalent to a system for one space dimension or to a system for one dependent function. The procedure used allowed also obtaining of the general solution for a coupled system of the eikonal and Hamilton-Jacobi equation.
\end{abstract}

\section{Background ideas}
We consider an overdetermined system of two eikonal equations for two functions and another equation linking these two functions
\begin{gather}
u_\mu u_\mu =0,\\ \nonumber
v_\mu v_\mu =0,\\ \label{eik2}
u_\mu v_\mu =1, \nonumber
\end{gather}
where $u=u(x_0,x_1,x_2)$,\ $v=v(x_0,x_1,x_2)$. In this section lower indices of dependent variables designate derivatives with respect to the relevant variables $x_\mu$, $\mu=0,1,2$ and summation is implied over the repeated indices. We assume that all functions considered are sufficiently smooth, and all dependent and independent variables take values in the real space. 

The system (\ref{eik2}) is a special case of a more general system 
\begin{gather}\label{h}
u_\mu u_\mu =0,\\ \nonumber
v_\mu v_\mu =0,\\ \nonumber
u_\mu v_\mu =h(u,v) \nonumber
\end{gather}
 \noindent 
with an arbitrary function $h(u,v)$. 

The system (\ref{h}) can be obtained as a result of local transformations of the system 
\begin{gather} \label{rho}
u_\mu u_\mu =\rho(u,v),\\ \nonumber
v_\mu v_\mu =\sigma(u,v),\\ \nonumber
u_\mu v_\mu =\tau(u,v) \nonumber
\end{gather}
\noindent 
with arbitrary functions $\rho$, $\sigma$ and $\tau$ when $\rho \sigma - \tau^2<0$ 
that appeared in \cite{cypr08} as a part of the reduction conditions of the multidimensional nonlinear wave equation $\Box \phi=F(\phi)$ by means of an anzatz with two new independent variables 
$\phi=\phi(\omega_1,\omega_2)$.

The general form of the system of the type (\ref{rho}) which can be reduced to the system (\ref{eik2}) under consideration is
\begin{gather*}
u_\mu u_\mu =2A_a(a,b)A_b(a,b),\\ 
v_\mu v_\mu =2B_a(a,b)B_b(a,b)),\\ 
u_\mu v_\mu =A_a(a,b)B_b(a,b)+B_a(a,b)A_b(a,b).
\end{gather*}
\noindent
where $a=a(u,v)$, $b=b(u,v)$ are arbitrary sufficiently smooth functions. 

An example of such system is 
$$
u_\mu u_\mu =1, \quad v_\mu v_\mu = -1, \quad u_\mu v_\mu =0. 
$$ 

Systems of coupled eikonal equations are interesting not only in the context of reduction of multidimensional equations. They appear in geometric optics, imaging, continuum mechanics etc. Numerous papers on numerical methods for such systems imply that finding of their general solutions may be interesting at least for testing of   these methods.

The system for two space variables is the simplest system of the form (\ref{eik2}) for which it is possible to obtain non-trivial solutions. In the case of one space variable we will have only a linear solution
$u=a(x_0 \pm x_1)+c_1$, $v=1/2a(x_0 \mp x_1)+c_2$ where $a={\rm const} \neq 0$, $c_1$ and $c_2$ are arbitrary constants.
 
We found a parametric general solution for the system (\ref{eik2}) and two space variables (the general solution here means to exclude special cases)
\begin{gather*}
u=\frac{x_1+\frac{x_2 z}{\sqrt{1-z^2}}-k'(z)}{g'(z)},\\ 
v=\frac{gx_2}{\sqrt{1-z^2}}+
\frac{p(z)}{g'(z)}\left[x_1+\frac{x_2z}{\sqrt{1-z^2}}-k'(z)\right]+r(z), \\
0=x_0-x_1z+x_2\sqrt{1-z^2}+ 
\frac{g(z)}{g'(z)}\left(x_1+\frac{x_2 z}{\sqrt{1-z^2}}-k'(z)\right)-k(z). 
\end{gather*}

Here
\begin{gather*}
r'=-k''(zg+(1-z^2)g'),\\ 
p=\frac{1}{2}(-g'^2+(g-zg')^2).
\end{gather*}

In this solution we have only two independent arbitrary functions.
We did not consider any special cases in the process of finding of the general solution, e.g. flat wave solutions or when the system is reduced to the system for one dependent function. Special cases will be considered in further papers dedicated to investigation of the various classes of solutions. 

This paper is intended as a technical paper to present a very detailed procedure for finding of the general solution by application of the hodograph and contact transformations.  

The method we use here was developed on the basis of ideas presented in the papers by Zhdanov, Revenko and Fushchych \cite{FZhR-preprint}, \cite{FZhR-JMPh} on the general solution of the d'Alembert-Hamilton system.

\section{Application of hodograph and contact transformations}
\subsection{Formulae for hodograph transformations}

We start from the functions
\begin{gather}\label{uv}
u=u(x_0,x_1,x_2), \qquad  v=v(x_0,x_1,x_2) 
\end{gather}
\noindent 
and assume that $u_{x_0} \neq 0$, as
otherwise the first equation of (\ref{eik2}) will have only constant solutions.

We go from the set (\ref{uv}) to new dependent variables $w$ and $v$, and new independent variables $y_0$, $y_1$, $y_2$.
\begin{equation} \label{hd}
u=y_0, \qquad x_0=w,\qquad  x_1=y_1,\qquad x_2=y_2.
\end{equation}

Expressions for derivatives
\begin{gather} \label{hderiv}
u_{x_0}=\frac{1}{w_{y_0}},\qquad
u_{x_1}=-\frac{w_{y_1}}{w_{y_0}},\qquad
u_{x_2}=-\frac{w_{y_2}}{w_{y_0}}, \\
v_{x_0}=\frac{v_{y_0}}{w_{y_0}},\qquad
v_{x_1}=v_{y_1}-v_{y_0}\frac{w_{y_1}}{w_{y_0}},\qquad
v_{x_2}=v_{y_2}-v_{y_0}\frac{w_{y_2}}{w_{y_0}}. \nonumber
\end{gather}

\subsection{Hodograph transformation applied}

Please note that in these new equations we denote derivatives with respect to $y_\mu$ as $v_{y_\mu}$ and $w_{y_\mu}$.

Substitution of the formulae for derivatives (\ref{hderiv}) to the first equation of (\ref{eik2}) gives
\begin{gather*}
-\frac{w_{y_1}^2}{w_{y_0}^2}-\frac{w_{y_2}^2}{w_{y_0}^2}+\frac{1}{w_{y_0}^2}=0. 
\end{gather*}

As we assumed $w_{y_0}\neq 0$, this equation is equivalent to
\begin{equation*} 
w_{y_1}^2+w_{y_2}^2=1.\label{eik3.1}
\end{equation*}

Substitution to the second equation of (\ref{eik2}) gives
\begin{gather*}
v_{y_1}^2+v_{y_0}^2 \frac{w_{y_1}^2}{w_{y_0}^2}
-2\frac{v_{y_0} v_{y_1} w_{y_1}}{w_{y_0}}+ v_{y_2}^2+
v_{y_0}^2 \frac{w_{y_2}^2}{w_{y_0}^2}-2\frac{v_{y_0}v_{y_2}w_{y_2}}{w_{y_0}}= \frac{v_{y_0}^2}{w_{y_0}^2},
\end{gather*}
\noindent
then
\begin{gather}\label{ee}
v_{y_1}^2+ v_{y_2}^2-2(v_{y_1} w_{y_1}+v_{y_2} w_2)\frac{v_{y_0}}{w_{y_0}}= 0.
\end{gather}

Substitution to the third equation of (\ref{eik2}) gives
\begin{gather*}
\frac{v_{y_0}}{w_{y_0}^2}+\frac{w_{y_1}}{w_{y_0}}\left(v_{y_1}+v_{y_0}\frac{w_{y_1}}{w_{y_0}}\right) +\frac{w_{y_2}}{w_{y_0}}\left(v_{y_2}+v_{y_0}\frac{w_{y_2}}{w_{y_0}}\right)=1.
\end{gather*}

As $w_{y_0}\neq 0$, we can have
\begin{gather*}
v_{y_1} w_{y_1}+v_{y_2} w_{y_2}=w_{y_0}.
\end{gather*}

Then the equation (\ref{ee}) takes the form
\begin{equation*} 
v_{y_1}^2+v_{y_2}^2=2v_{y_0}.
\end{equation*}

The resulting system of the transformed equation will be
\begin{gather} 
w_{y_1}^2+w_{y_2}^2=1, \nonumber \\
v_{y_1}^2+v_{y_2}^2=2v_{y_0},  \label{eik4} \\
v_{y_1} w_{y_1} +v_{y_2} w_{y_2}=w_{y_0}. \nonumber
\end{gather}

Note that we obtained a system that includes an eikonal equation and a Hamilton-Jacobi equation similar to the reduction conditions for the Schr\"odinger
equation considered in \cite{anzSch}.

\subsection{Contact transformations}
New independent variables
$z_0=y_0$, $z_1=w_{y_1}$, $z_2=y_2$.

New dependent variables are
\begin{equation} \label{Hdef}
H(z_0,z_1,z_2) =y_1w_{y_1}-w, \qquad v=v(z_0,z_1,z_2).
\end{equation}
Relations for derivatives with respect to new independent
variables
\begin{gather} 
H_{z_0}=-w_{y_0}, \qquad H_{z_1}=y_1, \qquad H_{z_2}=-w_{y_2},\nonumber \\ 
v_{y_0}=v_{z_0}+v_{z_1}w_{y_0 y_1},\qquad
v_{y_1}=v_{z_1}+w_{y_1 y_1},\qquad
v_{y_2}=v_{z_2}+v_{z_1}w_{y_1 y_2},\label{Hderiv} \\ 
w_{y_1 y_1}=\frac{1}{H_{z_1 z_1}},\qquad
w_{y_1 y_2}=-\frac{H_{z_1 z_2}}{H_{z_1 z_1}},\qquad
w_{y_0 y_1}=-\frac{H_{z_0 z_1}}{H_{z_1 z_1}},\nonumber \\ 
w_{y_0 y_2}=-\frac{\left|
  \begin{matrix}
    H_{z_1 z_1} & H_{z_1 z_2} \nonumber \\ 
    H_{z_0 z_1} & H_{z_0 z_2} \nonumber \\ 
  \end{matrix}
\right|}{H_{z_1 z_1}}.
\end{gather}

\subsection{Substitution of contact transformations into system (\ref{eik4}) }
After the substitution the system (\ref{eik4}) results in
\begin{equation} \label{c1}
z_1^2+H_{z_2}^2=1,
\end{equation}

\begin{gather} \label{c2}
\left(\frac{v_{z_1}}{H_{z_1 z_1}}\right)^2+
\left(v_{z_2}-v_{z_1}\frac{H_{z_1 z_2}}{H_{z_1 z_1}}\right)^2=
2\left(v_{z_0}-v_{z_1}\frac{H_{z_1 z_2}}{H_{z_1 z_1}}\right),
\end{gather}
\begin{gather}\label{c3}
v_{z_1}\frac{z_1}{H_{z_1 z_1}}
-H_{z_2}\left(v_{z_2}-v_{z_1}\frac{H_{z_1 z_2}}{H_{z_1 z_1}}\right)=-H_{z_0}.
\end{gather}

The equation (\ref{c1}) has a general solution for the function $H$
\begin{equation} \label{H}
H=z_2\sqrt{1-z_1^2}+G(z_0,z_1),
\end{equation}
where $G$ is a function of its arguments that is to be determined below. 

 The case 
$z_1=w_{y_1}=-\frac{u_{x_1}}{u_{x_0}}=\pm1$ is a special case we do not consider here.

From the expression for the function $H$ (\ref{H}) we get
\begin{gather} 
H_{z_0}=G_{z_0}, \qquad
H_{z_1}=-\frac{z_1 z_2}{\sqrt{1-z_1^2}}, \qquad
H_{z_2}=\sqrt{1-z_1^2}, \nonumber \\
H_{z_0 z_1}=G_{z_0z_1}, \qquad
H_{z_1 z_2}=-\frac{z_1}{\sqrt{1-z_1^2}},  \label{H'}\\
H_{z_1 z_1}=-\frac{z_2}{\sqrt{1-z_1^2}}-
\frac{z_1^2z_2}{(1-z_1^2)^{\frac{3}{2}}} + G_{z_1 z_1}=
-\frac{z_2}{(1-z_1^2)^{\frac{3}{2}}} + G_{z_1 z_1}.\nonumber
\end{gather}

Then substitution of the contact transformations (\ref{H'}) and the expressions for the derivatives of the function $H$ into (\ref{c2}) gives
\begin{gather*}
v_{z_1}z_1+\left(G_{z_1 z_1}-\frac{z_2}{(1-z_1^2)^{\frac{3}{2}}}\right)
\left(G_{z_0}-v_{z_2}\sqrt{1-z_1^2}\right)+
v_{z_1}\sqrt{1-z_1^2})\left(-\frac{z_1}{\sqrt{1-z_1^2}}\right) \nonumber \\
\qquad {}= \left(G_{z_1 z_1}-\frac{z_2}{(1-z_1^2)^{\frac{3}{2}}}\right)
\left(G_{z_0}-v_{z_2}\sqrt{1-z_1^2}\right)=0.
\end{gather*}

As
\begin{gather*}
G_{z_1 z_1}-\frac{z_2}{(1-z_1^2)^{\frac{3}{2}}}\neq 0,
\end{gather*} then
\begin{equation*}
G_{z_0}-v_{z_2}\sqrt{1-z_1^2}=0
\end{equation*}
that gives an expression for $v$
\begin{equation}\label{v1}
v=\frac{G_{z_0}z_2}{\sqrt{1-z_1^2}}+P(z_0,z_1),
\end{equation}
where $P(z_0,z_1)$ is some function of its arguments to be determined further.

From (\ref{v1}) we get expressions for the derivatives of the function $v$:
\begin{gather}\label{v'}
v_{z_0}=\frac{G_{z_0z_0}z_2}{\sqrt{1-z_1^2}}+P_{z_0}, \nonumber \\
v_{z_1}=\frac{G_{z_0z_1}z_2}{\sqrt{1-z_1^2}}+
\frac{G_{z_0}z_1z_2}{(1-z_1^2)^{\frac{3}{2}}}+P_{z_1},\\
v_{z_2}=\frac{G_{z_0}}{\sqrt{1-z_1^2}}.\nonumber
\end{gather}

Substitution of (\ref{H'}), (\ref{v1}),(\ref{v'})  into (\ref{c2}) gives
\begin{gather}
v_{z_1}^2 + (v_{z_2}H_{z_1 z_1}-v_{z_1}H_{z_1 z_2})^2
=2H_{z_1 z_1}(v_{z_0}H_{z_1 z_1}-v_{z_1}H_{z_0 z_1}),\nonumber
\\
\left(\frac{G_{z_0z_1}z_2}{\sqrt{1-z_1^2}}+
\frac{G_{z_0}z_1z_2}{(1-z_1^2)^{\frac{3}{2}}}P_{z_1}\right)^2 +\nonumber\\
\left(\frac{G_{z_0}}{\sqrt{1-z_1^2}}\left(G_{z_1 z_1}-\frac{z_2}{(1-z_1^2)^{\frac{3}{2}}}\right)+
\frac{z_1}{\sqrt{1-z_1^2}}\left(P_{z_1}+\frac{G_{z_0z_1}z_2}{\sqrt{1-z_1^2}}+
\frac{G_{z_0}z_1z_2}{(1-z_1^2)^{\frac{3}{2}}}\right)\right)^2= \nonumber\\
= 2\left(G_{z_1 z_1}-\frac{z_2}{(1-z_1^2)^{\frac{3}{2}}}\right)\times\nonumber\\
\times\left(\left(G_{z_1 z_1}-\frac{z_2}{(1-z_1^2)^{\frac{3}{2}}}\right)
\left(\frac{G_{z_0z_0}z_2}{\sqrt{1-z_1^2}}+P_{z_0}\right)
-G_{z_0z_1}\left(\frac{G_{z_0z_1}z_2}{\sqrt{1-z_1^2}}+
\frac{G_{z_0}z_1z_2}{(1-z_1^2)^{\frac{3}{2}}}+P_{z_1}\right)\right).\label{eik1cond-z2^3}
\end{gather}

Further we can decompose these expressions by powers of $z_2$.

At $z_2^3$ we get $-2\frac{G_{z_0z_0}}{(1-z_1^2)^{\frac{7}{2}}}=0$, whence  $G_{z_0z_0}=0$.

At $z_2^2$ we get
\begin{gather*}
\left(\frac{G_{z_0z_1}}{\sqrt{1-z_1^2}}+
\frac{G_{z_0}z_1}{(1-z_1^2)^{\frac{3}{2}}}\right)^2 +
\left( - \frac{G_{z_0}}{(1-z_1^2)^2}+
\frac{G_{z_0z_1}z_1}{1-z_1^2}+\frac{G_{z_0}z_1^2}{(1-z_1^2)^2}\right)^2\nonumber \\
\qquad{}
= -\frac{2}{(1-z_1^2)^{\frac{3}{2}}}\left( - \frac{P_{z_0}}{(1-z_1^2)^{\frac{3}{2}}}
-G_{z_0z_1}\left(\frac{G_{z_0z_1}}{\sqrt{1-z_1^2}}+
\frac{G_{z_0}z_1^2}{(1-z_1^2)^{\frac{3}{2}}}\right)\right).
\end{gather*}
\begin{gather*}
G_{z_0z_1}^2(1-z_1^2)^2+2G_{z_0z_1}G_{z_0}z_1(1-z_1^2)+G_{z_0}^2 z_1^2+
(1-z_1^2)(z_1^2 G_{z_0z_1}^2-2z_1G_{z_0z_1}G_{z_0}+\nonumber \\
G_{z_0}^2)=  2(P_{z_0}+G_{z_0z_1}^2(1-z_1^2)+G_{z_0z_1}G_{z_0}z_1)
\end{gather*}
\begin{gather}
(G_{z_0}-z_1G_{z_0z_1})^2=2P_{z_0}+G_{z_0z_1}^2 \label{eik1cond-z2sq}
\end{gather}

At $z_2$ we get
\begin{gather*}
P_{z_1}
\left(\frac{G_{z_0z_1}}{\sqrt{1-z_1^2}}+
\frac{G_{z_0}z_1}{(1-z_1^2)^{\frac{3}{2}}}\right)+
\left(\frac{G_{z_0}G_{z_1z_1}+z_1P_{z_1}}{\sqrt{1-z_1^2}}\right)\times
\left(-\frac{G_{z_0}}{(1-z_1^2)^2}+
\frac{G_{z_0z_1}z_1}{1-z_1^2}+\frac{G_{z_0}z_1^2}{(1-z_1^2)^2}\right)\nonumber\\
\qquad{}
= G_{z_1z_1}\left(-\frac{P_{z_0}+G_{z_0z_1}G_{z_0}z_1}{(1-z_1^2)^{\frac{3}{2}}}-
\frac{G_{z_0z_1}^2}{\sqrt{1-z_1^2}}\right)-
\frac{P_{z_0}G_{z_1z_1}-G_{z_0z_1}P_{z_1}}{(1-z_1^2)^{\frac{3}{2}}}
\end{gather*}

\begin{gather*}
P_{z_1}(G_{z_0z_1}(1-z_1^2)+G_{z_0}z_1)+(G_{z_0}G_{z_1z_1}+ z_1P_{z_1})(G_{z_0z_1}z_1- G_{z_0})+\nonumber\\
G_{z_1z_1}G_{z_0z_1}^2(1-z_1^2)+G_{z_1z_1}(P_{z_0}+G_{z_0z_1}G_{z_0}z_1)+
P_{z_0}G_{z_1z_1}-G_{z_0z_1}P_{z_1}=0
\end{gather*}

\begin{gather}
G_{z_1z_1}(G_{z_0}(G_{z_0z_1}z_1- G_{z_0})+G_{z_0z_1}^2(1-z_1^2)+
2P_{z_0}+G_{z_0z_1}G_{z_0}z_1)=0 \label{eik1cond-z2}
\end{gather}

From the condition (\ref{eik1cond-z2}) we get
\begin{gather}
G_{z_1z_1}[2P_{z_0}+G_{z_0z_1}^2-(G_{z_0}- G_{z_0z_1}z_1)^2]=0. \label{eik1cond-z2A}
\end{gather}

The expression in square brackets of (\ref{eik1cond-z2A}) being equal to zero is equivalent to the condition obtained as a result
of gathering of coefficients at $z_2^2$  (\ref{eik1cond-z2sq}). So, we get no new conditions from coefficients at $z_2$.

At $z_2^0$ we get
\begin{gather*}
P_{z_1}^2+\frac{(G_{z_0}G_{z_1z_1}+z_1P_{z_1})^2}{1-z_1^2}=
2G_{z_1z_1}(P_{z_0}G_{z_1z_1}-G_{z_0z_1}P_{z_1})
\end{gather*}

\begin{gather*}
P_{z_1}^2(1-z_1^2)+(G_{z_0}^2G_{z_1z_1}^2+2z_1P_{z_1}G_{z_0}G_{z_1z_1}
+z_1^2P_{z_1}^2= \nonumber\\
=2G_{z_1z_1}(P_{z_0}G_{z_1z_1}-G_{z_0z_1}P_{z_1})-2z_1^2(P_{z_0}G_{z_1z_1}^2-
G_{z_0z_1}G_{z_1z_1}P_{z_1})
\end{gather*}

\begin{gather}
P_{z_1}^2 +G_{z_0}^2G_{z_1z_1}^2+2z_1P_{z_1}G_{z_0}G_{z_1z_1}=
2(1-z_1^2)G_{z_1z_1}(P_{z_0}G_{z_1z_1}- P_{z_1}G_{z_0z_1}) \label{eik1cond-z2^0}
\end{gather}

From (\ref{eik1cond-z2^3})  $G_{z_0z_0}=0$, from (\ref{eik1cond-z2sq}) we get that $P_{z_0z_0}=0$. So $G$ and $P$
have the following form
\begin{gather}
G=g(z_1)z_0+k(z_1)\label{GP1} \\
P=p(z_1)z_0+r(z_1) \nonumber
\end{gather}
where $g,k,p,r$ are some functions on $z_1$, conditions on which will be obtained below.

Substituting (\ref{GP1}) into (\ref{eik1cond-z2sq}) we get
\begin{equation}
2p+g'^2=( g-z_1g')^2 \label{pg1}
\end{equation}

Substituting (\ref{GP1}) into (\ref{eik1cond-z2^0}) we get
\begin{gather}
(p'z_0+r')^2+g^2(g''z_0+k'')^2+2z_1(p'z_0+r')g(g''z_0+k'')=\nonumber \\
2(1-z_1^2)(g''z_0+k'')(p(g''z_0+k'')-g'(p'z_0+r'))
\end{gather}

Further we group coefficients at powers of $z_0$. At $z_0^2$ we get
\begin{gather}
p'^2+g^2g''^2+2z_1p'gg''=2(1-z_1^2)(g''^2p-g''g'p') \label{eik1cond-z0^2}
\end{gather}

From (\ref{pg1}) we come at the expression for the function $p$ from $g$:
\begin{gather}
p=\frac{1}{2}(g^2-2z_1gg'+(z_1^2-1)g'^2) \label{eik1cond-p1}
\end{gather}

Whence
\begin{gather}
p'=g''((z_1^2-1)g'-z_1g) \label{eik1cond-p'}
\end{gather}

Substituting the above expressions for $p$ and $p'$ into (\ref{eik1cond-z0^2}),
we get that
\begin{gather}
g''^2[((z_1^2-1)g'-z_1g)^2+g^2g''^2+2z_1g((z_1^2-1)g'-z_1g)-\nonumber \\
(1-z_1^2)((g^2-2z_1gg'+(z_1^2-1)g'^2)-g'((z_1^2-1)g'-z_1g))]=0 \label{eik1cond-z0^2A} \end{gather}

In square brackets of (\ref{eik1cond-z0^2A}) we get identical zero, so new conditions on functions $G$ and $P$ are obtained from coefficients at $z_0^2$.

Grouping coefficients at $z_0$ we come to the following condition:
\begin{gather*}
2p'r'+2g^2g''k''+2z_1g(p'k''+r'g'')=2(1-z_1^2)(g''(pk''-r'g')+k''(pg''-p'g'))
\end{gather*}

Substituting the expressions (\ref{eik1cond-p1}) and (\ref{eik1cond-p'}) for $p$ and $p'$, we get
\begin{gather*}
2g''[(r'+z_1gk'')((z_1^2-1)g'-z_1g)+(r'z_1g +k''g^2)]=\nonumber\\
2(1-z_1^2)g''[k''(g^2-2z_1gg'+(z_1^2-1)g'^2)-r'g'-k''g'((z_1^2-1)g'-z_1g)]
\end{gather*}
that results in an identity, so we once more get no new conditions.

At $z_0^0$ we get
\begin{gather}
r'^2+g^2k''^2+2z_1r'gk''-2(1-z_1^2)k''(pk''-g'r')= \nonumber\\
(r'-k''((z_1^2-1)g'-z_1g))^2=0,\label{eik1cond-z0^0}
\end{gather}

\noindent
and from (\ref{eik1cond-z0^0}) it follows that
\begin{gather*}
r'=k''((z_1^2-1)g'-z_1g),
\end{gather*}
\noindent
then if $g''\neq0$ it is equivalent to
\begin{gather*}
r'g''-p'k''=0.
\end{gather*}

Thus, we found the form of the functions $G$ and $P$
\begin{gather}
G=g(z_1)z_0+k(z_1)\nonumber \\
P=p(z_1)z_0+r(z_1),\label{GP2}
\end{gather}
\noindent
where
\begin{gather}
p=\frac{1}{2}(g^2-2z_1gg'+(z_1^2-1)g'^2)\nonumber \\
r'=k''((z_1^2-1)g'-z_1 g).\label{pr2}
\end{gather}

\section{Inverse contact and hodograph transformations}
We found the function $H$ as
\begin{equation*}
H=z_2\sqrt{1-z_1^2}+G(z_0,z_1),\label{H1}
\end{equation*}
\noindent
where $G(z_0,z_1)$ has the form (\ref{GP2}) with arbitrary $g$ and $k$,
and
\begin{equation*}
v=\frac{G_{z_0}z_2}{\sqrt{1-z_1^2}}+P(z_0,z_1),\label{v2}
\end{equation*}
\noindent
where $P(z_0,z_1)$ has the form (\ref{GP2}) with the functions $p$ and $r$ defined as (\ref{pr2})

The function $w$ can be determined from $H$ using transformations inverse to (\ref{hd}) and (\ref{hderiv}):
\begin{gather*}
w=z_1 H_{z_1}-H.
\end{gather*}

We can further relabel $z_1$ as $z$.
Thus we get a parametric general solution for the system (\ref{eik4})
\begin{gather*}
v=\frac{G_{z_0}z_2}{\sqrt{1-z^2}}+p(z)z_0+r(z), \\
w=y_1 z -y_2 \sqrt{1-z^2}-g(z)y_0 -k(z), \label{vw} \\
0=y_1+\frac{y_2z}{\sqrt{1-z^2}}-g'(z)y_0 -k'(z).\nonumber
\end{gather*}

Applying transformations inverse to (\ref{hd}) and (\ref{hderiv}), we get a parametric general solution of the coupled eikonal system (\ref{eik2}) for the original functions $u$ and $v$:
\begin{gather*}
u=\frac{x_1+\frac{x_2 z}{\sqrt{1-z^2}}-k'(z)}{g'(z)},
\\
v=\frac{gx_2}{\sqrt{1-z^2}}+
\frac{p(z)}{g'(z)}[x_1+\frac{x_2z}{\sqrt{1-z^2}}-k'(z)]+r(z),
\\
0=x_0-x_1z+x_2\sqrt{1-z^2}+
\frac{g(z)}{g'(z)}\{x_1+\frac{x_2 z}{\sqrt{1-z^2}}-k'(z)\}-k(z).
\end{gather*}

where
\begin{gather*}
r'=-k''(zg+(1-z^2)g'),\\
p=\frac{1}{2}(-g'^2+(g-zg')^2),
\end{gather*}
\noindent
$g$ and $k$ are arbitrary functions. Note that we skipped consideration of the special cases.

\section{Conclusions}
We presented a procedure that allowed construction of the general exact solution 
for the coupled system of eikonal equations. These results would allow e.g. describing of all ansatzes reducing an eikonal equation in two space variables to equations in one space variable, extending results on its symmetry reduction obtained e.g. in \cite{FSerovLdAE} and \cite{FSS}. 

Further research may include investigation and classification of the obtained solutions and generalization of the obtained results to higher dimensions.

\end{document}